\documentclass[twocolumn]{article}
\pdfoutput=1
\usepackage{amsthm,amsmath}
\usepackage[utf8]{inputenc} 
\usepackage[flushleft]{threeparttable}
\usepackage{upgreek}
\usepackage{amssymb}
\usepackage{url}
\usepackage{graphicx}
\usepackage[numbers]{natbib}

\newcommand{\Sint}{\mathbf{S_{int}}}

\title{Removing both Internal and Unrealistic Energy-Generating Cycles in Flux Balance Analysis}
\author{Elad Noor}
\begin{document}
\maketitle

\section{Abstract}
Constraint-based stoichiometric models are ubiquitous in metabolic research, with Flux Balance Analysis (FBA) being the most widely used method to describe metabolic phenotypes of cells growing in steady-state. Of the many variants of constrain-based modelling methods published throughout the years, only few have focused on thermodynamic issues, in particular the elimination of non-physical and non-physiological cyclic fluxes. In this work, we revisit two of these methods, namely thermodynamic FBA and loopless FBA, and analyze the strengths and weaknesses of each one. Finally, we suggest a compromise denoted \textit{semi-thermodynamic FBA} (st-FBA) which imposes stronger thermodynamic constrains on the flux polytope compared to loopless FBA, without requiring a large set of thermodynamic parameters as in the case of thermodynamic FBA. We show that st-FBA is a useful and simple way to eliminate thermodynamically infeasible cycles that generate ATP.

\section{Introduction}
The repertoire of genome-scale metabolic reconstructions is growing quickly, with more and more organisms being modeled every year \cite{King2016-it}. Likewise, the list of constraint-based modeling methods is getting longer with more than 100 papers published since 1961 in a rapidly increasing rate \cite{Lewis2012-mu} (see \url{http://cobramethods.wikidot.com/methods}). Flux Balance Analysis (FBA), is arguably the most well-known method to describe possible steady-state fluxes in a growing cell, and typically used to predict biomass yield and reaction essentiality. 

Throughout the years, many extensions have been suggested in order to incorporate different aspects of thermodynamics into constrain-based frameworks \cite{Beard2002-xt, Warren2007-wm, Henry2007-xp, Schellenberger2011-bq, Price2006-ua, Bordel2010-pl, Fleming2010-py, Holzhutter2004-qj, Fleming2009-um, Kummel2006-qn, Henry2006-nt, Hoppe2007-sw, De_Martino2012-cj, Tepper2013-gd, Nolan2006-eg, Nagrath2007-bn, Boghigian2010-vz}.
Two of these methods, \textit{thermodynamic FBA} \cite{Henry2007-xp} and \textit{loopless FBA} \cite{Schellenberger2011-bq} (also known as TMFA and ll-COBRA), define a new list of variables that represent the Gibbs free energy of formation of each metabolite in the system, and add constraints on reaction directionality that correspond to these free energies. Thermodynamic FBA imposes an extra set of constraints defining a range of possible values for each formation energy, and therefore requires many more parameters (which are often difficult to obtain). On the other hand, loopless FBA requires no extra parameters but does not necessarily eliminate all thermodynamically infeasible reactions or pathways. In this paper, we suggest a compromise denoted \textit{semi-thermodynamic FBA} (st-FBA) which imposes stronger thermodynamic constrains on the flux polytope compared to loopless FBA, without requiring a large set of thermodynamic parameters. Finally, we show that st-FBA is a useful and simple way to prevent flux in Energy Generating Cycles (EGCs) -- i.e. thermodynamically infeasible cycles that generate ATP or other energy currencies.

Typically, FBA requires a reconstruction of the metabolic network with $n$ metabolites and $r$ reactions, which is described by a stoichiometric matrix $\mathbf{S} \in \mathbb{R}^{n \times r}$. Some of the reactions in $\mathbf{S}$ are denoted \emph{primary exchange} reactions. These reactions represent the exchange of material between the model and the environment, i.e. input of nutrients and export of by-products. These reactions are not real chemical transformations nor transport reaction, but rather ``conceptual'' constructs that enable the system to be in a non-trivial steady state. Typically, primary exchange reactions have only one product and no substrates, or vice versa (see magenta reactions in Figure \ref{fig:cycles}). Another subset of reactions correspond to \emph{currency exchange} fluxes. These reactions represent the exchange of currency metabolites between different sub-networks (or compartments) in the cell. In the context of this work, we consider only the exchange of energy (e.g. ATP hydrolysis) as currency exchange. For instance, this exchange can be used to model the exchange of energy between the mitochondria (where ATP is generated) and the nucleus (where ATP is used). In bacterial models, where compartments play a minor role, many ATP utilizing processes are lumped into one cytoplasmic reaction called \emph{ATP maintenance} (see green reaction in Figure \ref{fig:cycles}). Finally, all other reactions are considered \emph{internal} reactions, and the sub-matrix of $\mathbf{S}$ corresponding to them is denoted $\Sint$.

\subsection{Steady-state assumption}
$\mathbf{S}$ relates between the flux vector $v \in \mathbb{R}^{r}$ and the change in metabolite concentrations $x \in \mathbb{R}^{n}$:
\begin{equation}
\frac{dx}{dt} = \mathbf{S} \cdot v\,\,.
\end{equation}
The steady-state assumption imposes a constraint that all concentrations are constant over time, therefore
\begin{equation}\label{eq:st1}
\mathbf{S} \cdot v = 0 \,\,. 
\end{equation}
Most constrain-based applications come with additional constraints on individual fluxes, i.e.
\begin{equation}\label{eq:st2}
\forall i~~~\alpha_i \leq v_i \leq \beta_i\,\,.
\end{equation}

\subsection{Extreme pathways}
Extreme pathways are convex basis vectors that define the polytope of solutions to the steady-state problem (Equations \ref{eq:st1}-\ref{eq:st2}). In \cite{Schilling2000-rn, Beard2002-xt}, three basic categories of extreme pathways were defined:
\begin{description}
\item[Type I] -- \textit{Primary systemic pathways}, i.e. pathways where at least one primary exchange flux is active.
\item[Type II] -- \textit{Futile cycles}, i.e. pathways where none of the primary exchange fluxes are active, but at least one currency exchange flux is active.
\item[Type III] -- \textit{Internal cycles}, i.e. pathways where none of the exchange fluxes (primary and currency) are active. 
\end{description}

\begin{figure}[ht!]
	\includegraphics[width=3in]{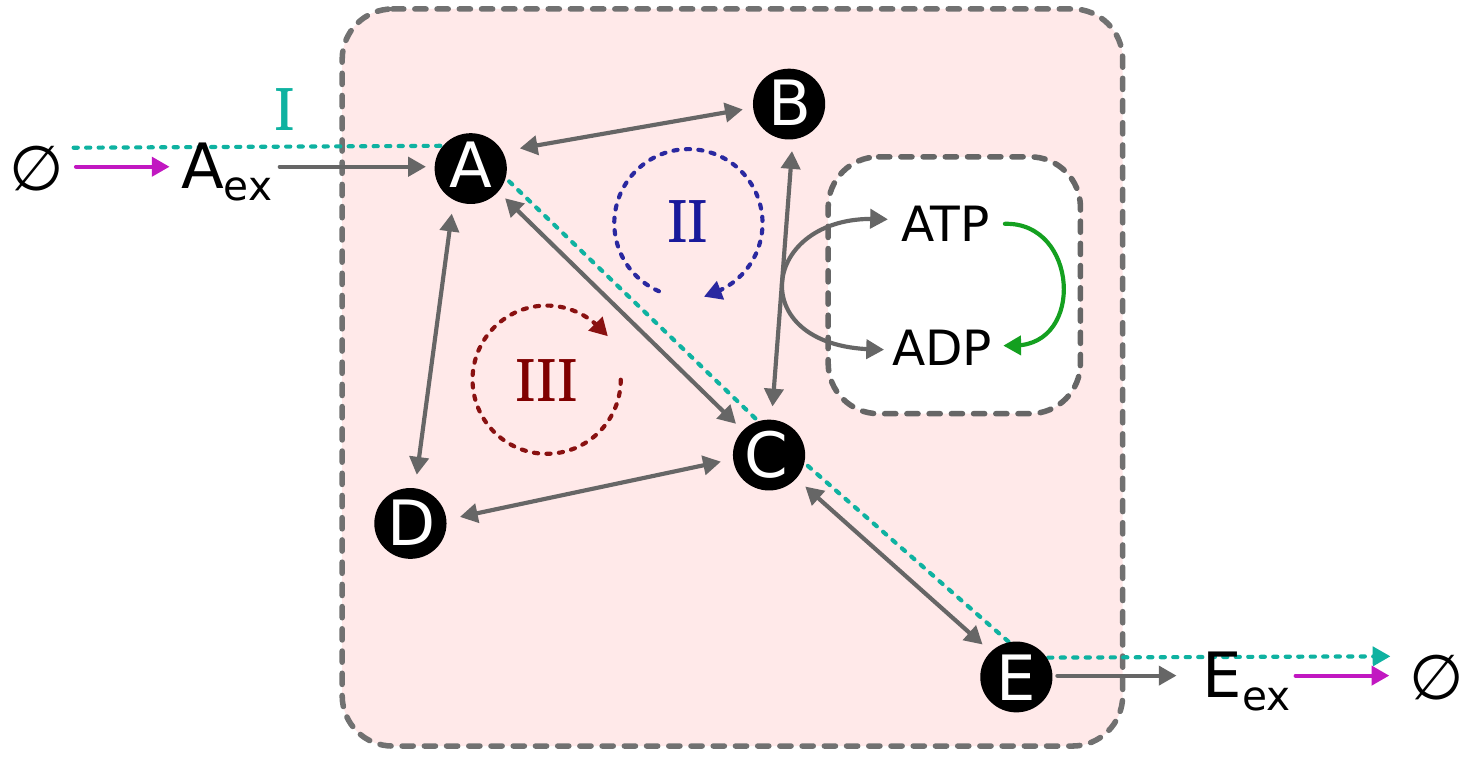}
	\caption{\textbf{The Three Types of Extreme Pathways.}
		A small toy model with 8 internal reactions (grey), 2 primary exchange reactions (magenta), and one currency exchange reaction (green). In this simple network, one can identify all three types of extreme pathways. The Type I pathway (connecting between $A_{ex}$ and $E_{ex}$, cyan) is typically the type of solution that most constrain-based models are seeking. The Type II pathway ($A \rightarrow B \rightarrow C \rightarrow A$, blue) is a typical futile cycle, since it does not involve any primary exchange reactions, but does \emph{waste} ATP. The Type III pathway ($A \rightarrow C \rightarrow D \rightarrow A$, red) is called internal since none of its reactions are exchange reactions. An internal cycle will never be thermodynamically feasible.}
    \label{fig:cycles}
\end{figure}
Figure \ref{fig:cycles} demonstrates this classification in a toy model. It is important to note, that this classification depends on the definition of primary and currency exchange reactions, and therefore is not a property of the stoichiometric matrix itself.

Many algorithms have been proposed for identifying and eliminating internal cycles (sometimes called "infeasible loops") from the solution space \cite{Price2002-ef, Kummel2006-qn, Price2006-ua, Wright2008-rh}. A more recent method denoted ``loopless'' FBA (or ll-FBA) \cite{Schellenberger2011-bq}, adds extra binary variables and constraints to the standard FBA framework, which effectively eliminate all type III pathways from the solution space, without removing any of the other solutions -- namely type I and type II pathways (as proven mathematically in \cite{Noor2012-qb}).
On its face, it seems to be exactly what one would want. Type I pathways are exactly the type of steady-state flux solutions one seeks in constraint-based models. Type II pathway might seem inefficient for the cell as they ``waste'' ATP without having any metabolic function, but they are still feasible and are even known to operate in vivo. Therefore, removing type II pathways from the solution space might impinge on the predictive value of a model.

Nevertheless, there are cases where type II pathways must be removed. In some cases, especially in automatically generated stoichiometric models \cite{Fritzemeier2017-ba}, one can find type II cycles that \textit{generate} ATP. A simple example would be a standard futile cycle running in reverse. This special case of type II pathways is denoted \textit{Energy Generating Cycle} (EGC). Typically, reaction directionality constraints are imposed to prevent such cycles, but it is often the case that some of these EGCs have been overlooked and are still possible flux solutions in published models. Fritzemeier et al. \cite{Fritzemeier2017-ba} found that this is the case in most network reconstructions in ModelSEED (\url{http://modelseed.org/}) and MetaNetX (\url{http://www.metanetx.org/}). 

Some might ask, why aren't EGCs eliminated by ll-FBA, as they are also thermodynamically infeasible cycles? To answer this, it is important to understand that type III cycles and EGCs are infeasible in two different ways. Internal type III cycles stand in violation of the first law of thermodynamics, i.e. conservation of energy states. A type III cycle is equivalent to a type one perpetual motion machine, or a river flowing in a complete circle\footnote{It would be more precise to say that type III cycles violate either the first or the second law. In essence, an internal cycle would not violate the first law if it were not generating any heat. However, in order to have a net flux in a biochemical reaction, the reaction must be out of equilibrium (according to the second law) and therefore the molecules flowing from the high energy state to the lower state would necessarily generate heat.}. EGCs are different, as they do not form a complete chemical cycle but are rather coupled to an ATP forming reaction. One could imagine a world where ADP + P$_i$ $\rightleftharpoons$ ATP + H$_2$O was a favorable reaction that could drive the other part of the cycle\footnote{Another way of looking at this, is to imagine running all the reactions in reverse. If we reverse all the reactions in a type III cycle, we will still have a type III cycle. If we do the same for an EGC, we would get a futile type II cycle, which is thermodynamically feasible.}. Therefore, EGCs do not violate the first law of thermodynamics, and are only infeasible in light of what we know about ADP and ATP in physiologically relevant conditions. In other words, EGCs violate the second law of thermodynamics, i.e. they altogether decrease the entropy of the universe. It is important to note here that type III cycles are often called Thermodynamically Infeasible Cycles (TICs) \cite{DeMartino2013, Desouki2015-lh}, but by now it should be clear that this can be confusing.

In the specific context of Flux Balance Analysis (FBA), EGCs are much more problematic than internal cycles, as their existence can increase the maximal yield of the metabolic network. A typical scenario would be an ATP-coupled cycle that effectively creates ATP from ADP and orthophosphate while all other intermediate compounds are mass balanced. This is equivalent to making ATP without any metabolic cost, which could effectively satisfy the ATP requirement of the biomass function and allow more resources to be diverted to biosynthesis. As we pointed out earlier, in well curated models such as the genome-scale \emph{E. coli} model \cite{Carrera2014-ys}, EGCs have been eliminated by manually constraining the directionality of many reactions (specifically, ATP coupled reactions). Although this is an effective way of removing EGCs, it has two major disadvantages: (i) it imposes hard constraints on reactions that might otherwise be reversible, and (ii) it is labor intensive and thus not scalable.

Recently, an automated method based on \textsc{GlobalFit} was shown to successfully eliminate almost all EGCs by removing a small number of reactions from the network \cite{Fritzemeier2017-ba}. This method makes it much easier to scale up to virtually all available metabolic network reconstructions, but does not deal with the first problem of over-constraining the flux solution set.

\subsection{Example of an Energy Generating Cycle in iJO1366}
\begin{figure*}[ht!]
	\includegraphics[width=6in]{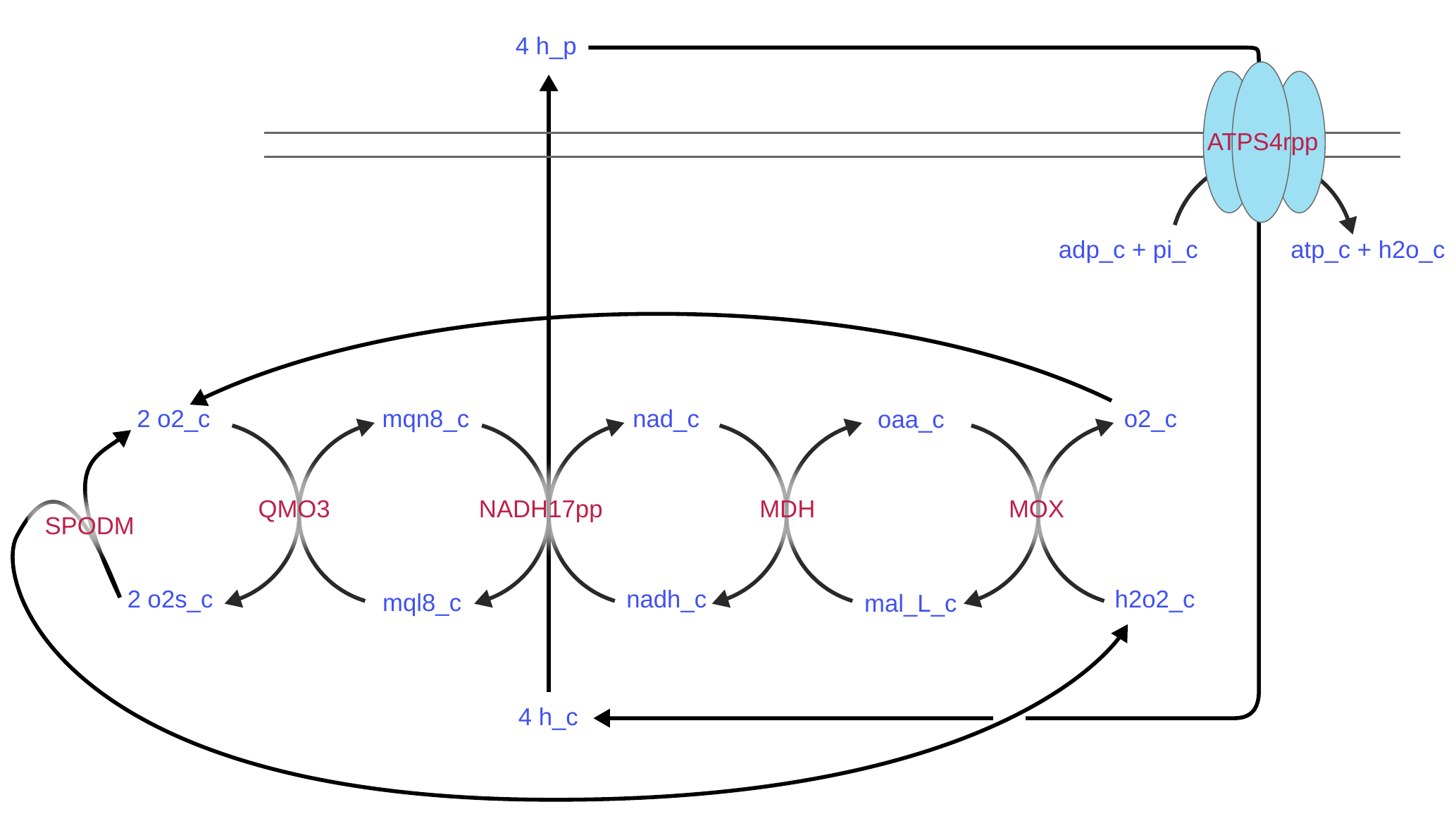}
 	\caption{\textbf{An Energy Generating Cycle.}
 		Reaction (red) and metabolite (blue) names are according to BiGG.
 		If the reactions SPODM and MOX are not removed from the model iJO1366, they can
 		be used in this unrealistic pathway that generates ATP without any
 		external input.}
 	\label{fig:egc_example}
\end{figure*}

One of the well-known examples for an EGC appears in the latest genome-scale reconstruction of \emph{E. coli} metabolism, denoted iJO1366 \cite{Orth2011-qi}. Orth et al. published the model together with a warning that ``hydrogen peroxide producing and consuming reactions carry flux in unrealistic energy generating loops'' and therefore these reactions are constrained by default to zero (see table \ref{table:egc_example} and figure \ref{fig:egc_example}). Ideally, we would like to keep these reactions as they might be useful (or even essential) for the cell in some environments.

\begin{table*}[ht!]
	\caption{Example of an Energy Generating Cycle that is possible in iJO1366}
	\begin{tabular}{l|l}
		\label{table:egc_example}
		Reaction & Formula\\\hline
		MDH & mal\_\_L-c + nad\_c = h\_c + nadh\_c + oaa\_c\\
		MOX	& h2o2\_c + oaa\_c = mal\_\_L\_c + o2\_c\\
		Htex & h\_e = h\_p\\
		EX\_h\_e & = h\_e\\
		SPODM & 2.0 h\_c + 2.0 o2s\_c = h2o2\_c + o2\_c\\
		NADH17pp & 4.0 h\_c + mqn8\_c + nadh\_c = mql8\_c + nad\_c + 3.0 h\_p\\
		QMO3 & mql8\_c + 2.0 o2\_c = 2.0 h\_c + mqn8\_c + 2.0 o2s\_c\\
		ATPS4rpp & adp\_c + pi\_c + 4.0 h\_p = atp\_c + 3.0 h\_c + h2o\_c\\\hline
		Total & adp\_c + pi\_c = atp\_c + h2o\_c
	\end{tabular}
\end{table*}

Malate oxidase (MOX) reaction is used in the direction of oxaloacetate reduction. As was suggested previously \cite{Fritzemeier2017-ba}, constraining the reaction to be irreversible only in the direction of malate oxidation (after all, its $\Delta_r G'^\circ$ is approximately $-100$ kJ/mol) would solve this problem and eliminate the EGC from the solution space.

Nevertheless, many EGCs cannot be solved by constraining the direction of one thermodynamically irreversible reaction. Before giving an example for such a case, we must first explain the notion of \emph{distributed thermodynamic bottlenecks} \cite{Mavrovouniotis1996-dq, Mavrovouniotis1993-zq}.

\subsection{Distributed Thermodynamic Bottlenecks}
The second law of thermodynamics, as applied to enzyme-catalyzed reactions, states that a reaction is feasible only if the $\Delta_r G'$ of the chemical transformation is negative. Typically, we use the formula
\begin{eqnarray}
	\Delta_r G' = \Delta_r G'^\circ + RT \cdot \sum_i \nu_i \ln(c_i)\,\,,
\end{eqnarray}
where $c_i$ is the concentration of reactant $i$ and $\nu_i$ is its stoichiometric coefficient (negative for substrates and positive for products) and $\Delta_r G'^\circ$ is the change in standard Gibbs free energy of this reactions. In matrix notation, for all reactions in the system:
\begin{eqnarray}
\mathbf{\Delta_r G'} = \mathbf{\Delta_r G'^\circ} + RT \cdot \mathbf{S}^\top \ln(\mathbf{c})\,\,.
\end{eqnarray}
In a series of papers from the 90's, Michael Mavrovouniotis developed methods for estimating these standard Gibbs energy changes \cite{Mavrovouniotis1990-wv, Mavrovouniotis1991-kf}, and for quantifying what he defined as thermodynamic bottlenecks \cite{Mavrovouniotis1993-zq, Mavrovouniotis1996-dq}. Mavrovouniotis noticed that the classic approach to reversibility, which is based on rather arbitrary thresholds imposed on every single reaction's $\Delta_r G'$, is insufficient. Moreover, since the vector of metabolite concentrations -- $\ln(\mathbf{c})$ -- is common to all reactions, the constraints on $\mathbf{\Delta_r G'}$ arising from the second law of thermodynamics can be coupled in a way that does not allow sets of reactions to be feasible together, even though each single reaction is feasible individually. Mavrovouniotis denoted such cases as \emph{distributed bottlenecks}.

In the context of this work, we claim that distributed bottlenecks can also form EGCs. When such case occurs, it is especially difficult to eliminate these EGCs from the set of feasible fluxes, as no single directionality flux constraint can be justified thermodynamically. Only the \emph{combined} activity of all the EGC reactions is thermodynamically infeasible. Moreover, these distributed EGCs are not as rare as one might think.

\subsubsection{Example of Distributed Energy Generating Cycle in the Central Metabolism of E. coli}
Consider the following three enzyme-catalyzed reactions:
\begin{itemize}
	\item Pyruvate kinase (PYK):\\ADP + PEP $\rightleftharpoons$ ATP + pyruvate
	\item PEP synthase (PPS):\\ATP + pyruvate + H$_2$O $\rightleftharpoons$ AMP + PEP + P$_i$
	\item Pyruvate-phosphate dikinase (PPDK):\\ATP + pyruvate + P$_i$ $\rightleftharpoons$ AMP + PEP + PP$_i$
	\item Adenylate kinase (ADK1):\\2 ADP $\rightleftharpoons$ ATP + AMP
\end{itemize}
In most if not all metabolic models that include PYK and PPS (or PPDK), these reactions are marked as irreversible. However, this annotation is not based on thermodynamic reversibility constraints, but rather from higher-level knowledge about the system. Pyruvate kinase (PYK, EC 2.7.1.40) is used exclusively in glycolysis and its activity is inhibited when gluconeogensis is required for growth. However, the reaction itself is reversible, as has been shown in vitro \cite{Lardy1945-ze}. In fact, the equilibrium constant has been measured to be as high as $1.5 \times 10^{-3}$ in some conditions \cite{Krimsky1959-mt}. The equilibrium constant of PEP synthase (PPS, EC 2.7.9.2) has not been directly measured in vitro, but a similar reaction catalyzed by pyruvate-phosphate dikinase (PPDK, EC 2.7.9.2) was found to be reversible with an equilibrium constant of $\sim10^{-3}$ at neutral pH, and as high as 0.5 at pH 8.39 \cite{Reeves1968-mq}. Combining the reaction of PPDK and the pyrophosphatase reaction PP$_i$ $\rightleftharpoons$ 2 P$_i$, would yield the PPS reaction \cite{De_Meis1982-pe}. The equilibrium constant of pyrophosphatase has been measured directly numerous times, and lies between 100 and 1000, depending on pH. Therefore, the equilibrium constant overall combined reaction (PEP synthase) is the product of both these values and is somewhere very close to 1. According to estimates done by eQuilibrator\footnote{http://equilibrator.weizmann.ac.il/} K'$_{\rm eq}$ $\approx$ 1.3 when the pH is set to 7.4 and the ionic strength is 0.1~M \cite{Flamholz2011}.

\begin{figure}[ht!]
	\includegraphics[width=3in]{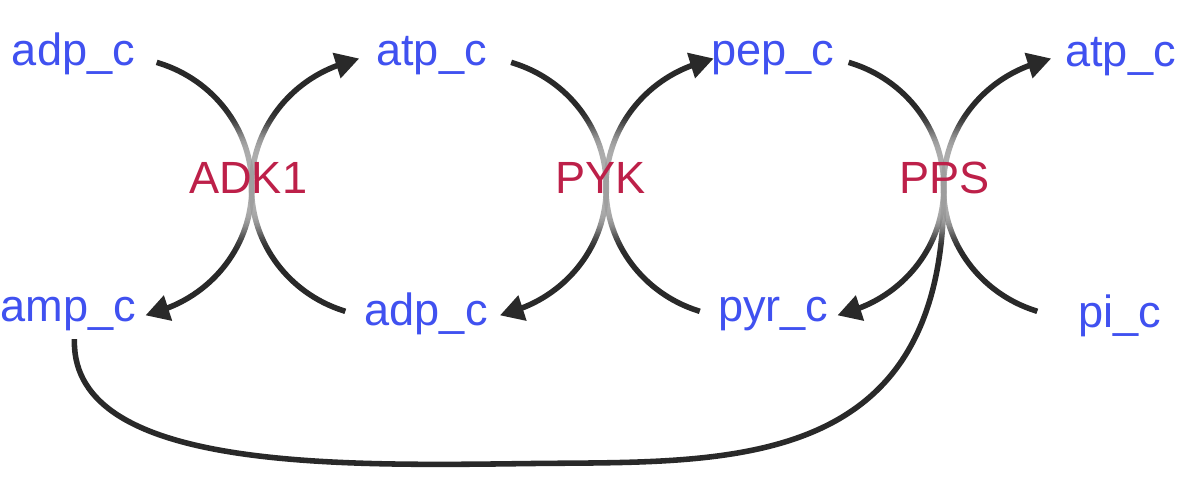}
	\caption{\textbf{An Energy Generating Cycle in \emph{E. coli} consisting of a distributed thermodynamic bottleneck.}
		One can see, that the overall reaction of the three combined enzymes gives ADP~+~P$_i$~$\rightleftharpoons$~ATP~+~H$_2$O}
	\label{fig:distributed_egc}
\end{figure}

Overall, we see that all three reactions (ADK1, PYK, and PPS) are individually reversible. Nevertheless, combining all three reactions can create an energy generating cycle, as shown in Figure \ref{fig:distributed_egc}. This is one of the simplest examples for distributed EGCs, and it appears in the most ubiquitous metabolic pathway of all, glycolysis. One might ask, then why doesn't this EGC cause problems for the iJO1366 model for \emph{E. coli}? There is a simple answer, PYK and PPS are both annotated as irreversible reactions in this model (and virtually all similar models). Admittedly, the regulatory network in \emph{E. coli} is hard-wired to prevent backward flux in PYK when PPS is active and vice versa, therefore constraining these fluxes in the model might not be very harmful. However, one can imagine scenarios where the reversibility of PPS and PYK might be important to consider. For example, in metabolic engineering projects, where the possibility of evolving bypasses to certain pathways could play a major role.

In the following sections, we describe two established methods for dealing with thermodynamic constraints and futile cycles in FBA models, discuss their strengths and weaknesses, and suggest a compromise that would specifically target distributed EGCs.

\subsection{Thermodynamic Flux Balance Analysis (TFBA)}

Thermodynamic FBA (also known as Thermo\-dynamic-based Metabolic Flux Analysis \cite{Henry2007-xp}) was designed to deal with thermodynamically infeasible flux solutions. However, its widespread adoption has been hampered by the requirement for thermodynamic parameters. The set of equations that describe TFBA are:
\begin{eqnarray}
\textsf{\textbf{TFBA}} && \nonumber\\
\mathbf{v^*} &=& \mathrm{arg\max_v} {~\mathbf{c}^\top\mathbf{v}}\nonumber\\
\textsf{such that:} && \nonumber\\
\mathbf{S} \cdot \mathbf{v} &=& \mathbf{0}  \label{eq:tfba1} \\
0 &\leq& M \mathbf{y} - \mathbf{v} ~\leq~ M
\label{eq:tfba2} \\
\varepsilon &\leq& M \mathbf{y} + \mathbf{\Delta_r G'} ~\leq~ M - \varepsilon \label{eq:tfba3} \\
\mathbf{v}^L &\leq& \mathbf{v} ~\leq~ \mathbf{v}^U \label{eq:tfba4}\\
\mathbf{y} &\in& \{0, 1\}^r \label{eq:tfba5}\\
\mathbf{\Delta_r G'} &=& \mathbf{\Delta_r G'^\circ} + RT \cdot \Sint^\top \mathbf{x} \label{eq:tfba6}\\
\ln(\mathbf{b}^L) &\leq& \mathbf{x} ~\leq~ \ln(\mathbf{b}^U) \label{eq:tfba7}
\end{eqnarray}
where $\mathbf{c} \in \mathcal{R}^r$ is the objective function, and the constants are the stoichiometric matrix of internal reactions $\Sint \in \mathcal{R}^{m \times r}$  and the vector of standard Gibbs energies of reaction $\mathbf{\Delta_r G'^\circ}$ (in units of kJ/mol), the gas constant $R$ = 8.31 J/mol/K and temperature $T$ = 300 K. It is important to verify that $\mathbf{\Delta_r G'^\circ}$ is orthogonal to the null-space of $\mathbf{S_{int}}$ (or, equivalently, in the image of $\mathbf{S_{int}}^\top$). Otherwise, this would be a violation of the first law of thermodynamics \cite{Noor2012-mp}.

The variables are the flux vector $\mathbf{v} \in \mathcal{R}_{+}^{r}$, the vector of binary reaction indicators $\mathbf{y} \in \{0,1\}^{r}$, and the vector of log-scale metabolite concentrations $\mathbf{x} \in \mathcal{R}^{m}$. $M$ is set to a positive number which is larger than any possible flux value and larger than any possible $\Delta_r G'$. $\varepsilon$ is a very small number (e.g. $10^{-9}$), usually added to LP constraints in order to achieve a strong inequality (for numerical stability reasons). Equation \ref{eq:tfba2} ensures that $\forall j~v_j > 0 \rightarrow y_j = 1$ and $v_j < 0 \rightarrow y_j = 0$, and equation \ref{eq:tfba3} ensures that $y = 1 \rightarrow \Delta_r G'_j < 0$ and $y = 0 \rightarrow \Delta_r G'_j > 0$. This forces any flux solution to follow the second law of thermodynamics \cite{Hoppe2007-sw, Machado2017-gh}, which is summarized as $\forall j:v_j = 0~\vee~\text{sign}(v_j) = -\text{sign}(\Delta_r G'_j)$. It is a rather different notation compared to the original paper \cite{Henry2007-xp}, where one had to assume all fluxes are positive (which usually requires making the model irreversible by decomposing each reversible reaction into two irreversible ones). Using the reversible notation for TFBA can reduce the number of boolean variable considerably and thus decrease the effective running time of the MILP solver.
Finally, Equations \ref{eq:tfba6}-\ref{eq:tfba7} together set the bounds on the Gibbs energies of reaction, assuming the concentration of each metabolite $i$ is between $b^L_i$ and $b^U_i$. Note that $\mathbf{\Delta_r G'}$ is not really a variable in the LP (as it is an affine transformation of $\mathbf{x}$), and is explicitly defined only for the sake of clarity. 

While the stoichiometric matrix ($\mathbf{S}$) and general flux constraints are exactly the same as in the standard FBA formulation, $\mathbf{\Delta_r G'^\circ}$ comes as an additional requirement for running TFBA. Unfortunately, we still lack precise measurement for many of the compounds comprising biochemical networks, and computational methods that estimate $\mathbf{\Delta_r G'^\circ}$ \cite{Jankowski2008-hd,Noor2012-mp,Noor2013-an,Jinich2014-nv} are far from perfect and sometimes introduce significant errors. The fact that TFBA also adds one boolean variable for each reaction in the model definitely doesn't help either, since solving the LP becomes much harder, requires a good MILP solver, and takes much longer than standard FBA. Due to the effort involved, and the unclear benefit of the method, TFBA has not gained a wide audience of users so far.

A more recent method called tEFMA \cite{Gerstl2015a, Gerstl2015} also aims to eliminate thermodynamically infeasible solutions, but in the context of Elementary Flux Mode Analysis (EFMA). Here, we focus only on FBA extensions, and therefore leave tEFMA and similar methods out of this comparison.

\subsection{Loopless Flux Balance Analysis (ll-FBA)}
In light of these caveats, it might be easier to understand why ll-FBA was introduced four years \emph{after} TFBA \cite{Schellenberger2011-bq}. Essentially, the loopless algorithm uses exactly the same MILP design as TFBA, while forgoing the actual thermodynamic values. This way, thermodynamically infeasible internal (Type III) cycles are eliminated, while all other pathways are kept \cite{Noor2012-qb}. The set of equations describing ll-FBA are:
\begin{eqnarray}
\textsf{\textbf{ll-FBA}} && \nonumber\\
\mathbf{v^*} &=& \mathrm{arg\max_v} {~\mathbf{c}^\top\mathbf{v}}\nonumber\\
\textsf{such that:} && \nonumber\\
\mathbf{S} \cdot \mathbf{v} &=& \mathbf{0} \label{eq:llfba1} \\
0 &\leq& M \mathbf{y} - \mathbf{v} ~\leq~ M
\label{eq:llfba2} \\
\varepsilon &\leq& M \mathbf{y} + \mathbf{\Delta_r G'} ~\leq~ M - \varepsilon \label{eq:llfba3} \\
\mathbf{v}^L &\leq& \mathbf{v} ~\leq~ \mathbf{v}^U \\
\mathbf{y} &\in& \{0, 1\}^r \label{eq:llfba4} \\
\mathbf{\Delta_r G'} &\in& (\ker{(\mathbf{S_{int}})})^\perp \label{eq:llfba5}
\end{eqnarray}
where all variables and constants are the same as in equations \ref{eq:tfba1}-\ref{eq:tfba7}. One change made here relative to the original formulation presented by Schellenberger et al. \cite{Schellenberger2011-bq}, is rewriting equations \ref{eq:llfba1}-\ref{eq:llfba2} to facilitate the comparison to TFBA. In addition, we use $\varepsilon$ rather than $1$ as the margin in equation \ref{eq:llfba2}.
Comparing this system of equations to TFBA (Equations \ref{eq:tfba1}-\ref{eq:tfba7}), one can easily see that they are identical except for the last two constraints (Equation \ref{eq:tfba6}-\ref{eq:tfba7}). Furthermore, Equation \ref{eq:llfba5} can actually be rewritten as $\mathbf{\Delta_r G'} = \mathbf{S_{int}}^\top \cdot \mathbf{\Delta_f G'}$, where $\mathbf{\Delta_f G'} \in \mathcal{R}^{m}$ is an unconstrained vector (this is a general form for a vector in the image of $\mathbf{S_{int}}^\top$, and from the fundamental theorem of linear algebra it is orthogonal to $\ker{(\mathbf{S_{int}})}$). Since $\mathbf{\Delta_f G'} = \mathbf{\Delta_f G'^\circ} + RT \cdot \mathbf{x}$, having it unconstrained is equivalent to having $\mathbf{x}$ completely unconstrained. Therefore, ll-FBA is simply TFBA with infinite concentration bounds.

\paragraph{Limited adoption of ll-FBA} Although ll-FBA requires no extra parameters compared to FBA, and its implementation is streamlined as part of the COBRA toolbox, it has yet to become mainstream. A plausible explanation would be that there is an alternative method for eliminating internal cycles in FBA solutions, which does not require an MILP, and can be easily implemented: after applying additional constraints that  define the relevant solution space (e.g. realizing the maximal biomass yield, or keeping all exchange fluxes constant), find a solution with the minimum sum of absolute (or squared) fluxes \cite{Holzhutter2004-qj}. Implementations of this principle with slight variations have been presented under different names such as parsimonious FBA \cite{Lewis2010-rx, Schuetz2012-sv} or CycleFreeFlux \cite{Desouki2015-lh}. Nevertheless, such methods are not suitable for some applications of ll-FBA, such as loopless Flux Variability Analysis (ll-FVA).

\paragraph{Related methods and improvements}
Already in 2002, Beard et al. \cite{Beard2002-xt} introduced Energy Balance Analysis (EBA), a method for enforcing the laws of thermodynamics in FBA simulations. The additional constraints that EBA enforces are essentially identical to ll-FBA, except that nonlinear optimization is applied instead of the MILP formulation introduced in equations \ref{eq:llfba2}-\ref{eq:llfba3}. Unfortunately, nonlinear optimization software tends to be much less efficient than state-of-the-art MILP solvers, and therefore the use of EBA was limited to relatively small models. Nevertheless, the methodology behind EBA was developed further and became a way to learn about chemical potentials from pure stoichiometric data \cite{Beard2004, Warren2007-wm, Reznik2013}. Network-embedded thermodynamic (NET) analysis \cite{Kummel2006-px, Kummel2006-qn, Zamboni2008} is derived from the same basic constraints (we discuss it in more detail in the following section).

More recently, two different approaches for reducing the number of boolean variables in ll-FBA have been published: Fast sparse null-space pursuit (fast-SNP) \cite{Saa2016} and Localized Loopless Constraints (LLCs) \cite{Chan2017}. These method do not change the ll-FBA problem, but rather try to get rid of constraints and boolean variables that are not necessary for solving it. Therefore, ll-FBA can be solved for much larger networks, and without the compromises that come with the methods that use flux minimization. It is still unclear whether these approaches can be extended to TFBA as well.

That being said, the aim of this work is not to discuss issues of complexity and runtime, but rather the theoretical implications of thermodynamic and loop\-less constraints on the set of feasible flux solutions.

\subsection{Network-Embedded Thermodynamic (NET) analysis}
NET analysis \cite{Kummel2006-px} is a highly related method which applies the same directionality constraints as TFBA, but does not aim to predict flux distributions. Instead, it aims to explore the ranges of possible $\Delta_r G'$ values, and can check for thermodynamic inconsistencies in quantitative metabolomic data sets. The NET analysis optimization problem (adapted from \cite{Kummel2006-px} to fit the notation in this manuscript) is:
\begin{eqnarray}
\textsf{\textbf{NET}} && \nonumber\\
\forall k && \mathrm{\min/\max} ~~\Delta_r G'_k\nonumber\\
\textsf{such that:} && \nonumber\\
\forall  v_i > 0 && \Delta_r G'_i < 0 \label{eq:net1}\\
\forall  v_i < 0 && \Delta_r G'_i > 0 \label{eq:net2}\\
\mathbf{\Delta_r G'} &=& \mathbf{\Delta_r G'^\circ} + RT \cdot \Sint^\top \mathbf{x} \\
\ln(\mathbf{b}^L) &\leq& \mathbf{x} ~\leq~ \ln(\mathbf{b}^U)\,.
\end{eqnarray}
Note that here the second law of thermodynamics is laid out explicitly (equations \ref{eq:net1}-\ref{eq:net2}), and not given as a pair of constraints involving boolean variables (as in TFBA and ll-FBA). This is facilitated by the fact that all intracellular flux directions are a prerequisite for the NET analysis, and therefore these constraints are hard-coded in the linear optimization problem. The only free variables are the log-concentrations in the vector $\mathbf{x}$ (as before, $\mathbf{\Delta_r G'}$ is just an affine transformation of $\mathbf{x}$).

Therefore, this formulation of NET analysis is a linear optimization problem, and does not require an MILP solver. However, in a more generalized case (such as the one used in \cite{Kummel2006-px}), constraints can also be imposed on the total concentrations of a subset of metabolites (for example, the sum of all pentose-phosphates). This requirement reflects the reality of many mass-spectrometry datasets where only the total concentration of several mass-isomers can be measured reliably. The downside is that these constraints are non-linear (due to the fact that our variables are log-concentrations, so the constraints would be on the sum of their exponents). 

NET analysis is akin to Flux Variability Analysis (FVA), except that the fluxes are fixed and the studied variability is the possible range of Gibbs free energies. A GUI-based Matlab\texttrademark~program called anNET \cite{Zamboni2008} facilitates the application of NET analysis in labs working on quantitative metabolomics.

\section{Results}

\subsection{A compromise between ll-FBA and TFBA}
Is there a version of thermodynamic-based FBA, that doesn't require a large set of extra (unknown) parameters, and still has a clear benefit over the standard tools that ignore thermodynamics? Here, we propose such a compromise, by relaxing the majority of second-law constraints (i.e. Equation \ref{eq:tfba3}) and keeping only a few important ones. We will show that this method, which we denote st-FBA (semi-thermodynamic Flux Balance Analysis), is sufficient to eliminate energy generating cycles, while requiring a relatively small set of heuristic assumptions and thermodynamic constants.

\begin{table*}[ht!]
	\caption{Table of currency metabolites, their standard Gibbs energies of formation, and bounded by their typical concentrations \cite{Bennett2009-rm}.}
	\begin{tabular}{l|c|c|r}
		\label{table:potentials}
		\textbf{metabolite} & $\mathbf{B_{low}}$ & $\mathbf{B_{high}}$ & $\Delta_f G'^\circ$ [kJ/mol] \\ \hline
		ATP & $9.63$ mM & $9.63$ mM & $-2296$ \\
		ADP & $0.56$ mM & $0.56$ mM & $-1424$ \\
		AMP & $0.28$ mM & $0.28$ mM & $-549$ \\
		orthophosphate & $1$ mM & $10$ mM & $-1056$ \\
		pyrophosphate & $1$ $\upmu$M & $1$ mM & $-1939$ \\
		H$^+$ (cytoplasm) & $10^{-7.6}$ M $^\dagger$ & $10^{-7.6}$ M & $0$ \\
		H$^+$ (extracellular) & $10^{-7.0}$ M & $10^{-7.0}$ M & $0$ \\
		H$_2$O & 1 & 1 & $-157.6$
	\end{tabular}
	\begin{tablenotes}
		\tiny
		\item $\dagger$ Corresponding to pH 7.6, as measured by \cite{Wilks2007-lh}
	\end{tablenotes}
\end{table*}

First, one must define a set of energy currency metabolites. Although the definition is somewhat heuristic, most biologists would agree that the following are energy equivalents: ATP, pyrophosphate, and a gradient of protons across the membrane. Other specific energy-carrying currency metabolites can be added to the list if desired. Next, we must bound the chemical potential ($\mathbf{\Delta_f G'}$) of these currency metabolites and all their associated degraded forms (see Table \ref{table:potentials}). Note that we chose the chemical potential at 1 mM concentration in an aqueous solution, which is a typical concentration for co-factors in \emph{E. coli} \cite{Bennett2009-rm}. Although it is possible to use the exact measured concentration of each of these metabolites, the effect on the st-FBA results would be at most very minor. 

Finally, we fix the values in the $\mathbf{G_f}$ vector only for these metabolites from the table, while the rest of the values remain free.
\begin{eqnarray}
\textsf{\textbf{st-FBA}}\nonumber\\
\mathbf{v^*} &=& \mathrm{arg\max_v} {~\mathbf{c}^\top\mathbf{v}}\nonumber\\
\textsf{such that:} && \nonumber\\
\mathbf{S} \cdot \mathbf{v} &=& \mathbf{0} \label{eq:stfba1}\\
0 &\leq& M \mathbf{y} - \mathbf{v} ~\leq~ M \label{eq:stfba2}\\
\varepsilon &\leq& M \mathbf{y} + \mathbf{\Delta_r G'} ~\leq~ M - \varepsilon\label{eq:stfba3}\\
\mathbf{v}^L &\leq& \mathbf{v} ~\leq~ \mathbf{v}^U \label{eq:stfba4}\\
\mathbf{y} &\in& \{0, 1\}^r \label{eq:stfba5}\\
\mathbf{\Delta_r G'} &=& \Sint^\top \cdot \mathbf{\Delta_f G'}\\
\mathbf{\Delta_f G'} &\in& \mathcal{R}^{m}\\
\forall i~\textsf{in Table \ref{table:potentials}}\nonumber\\
\Delta_f G'_i &\geq& \Delta_f G'^\circ_i + RT \cdot \ln(b^L_i) \\
\Delta_f G'_i &\leq& \Delta_f G'^\circ_i + RT \cdot \ln(b^U_i) 
\end{eqnarray}
So st-FBA is very similar to TFBA, except that only the energy currency metabolites have predefined bounds on their formation energies. In fact, since the concentrations of these metabolites tend to be tightly controlled by homeostasis, it is recommended to set them to fixed concentrations (i.e. by setting the lower and upper bounds to the same value). All other metabolites, on the other hand, have no constraints on their $\Delta_f G'$ (or, equivalently, no constraints on their concentrations) -- similar to the case in ll-FBA. In other words, $\mathbf{\Delta_r G'}$ is still a vector from $\ker{(\mathbf{S_{int}})}^\perp$, with a few extra constraints, but not as many as in TFBA.

\subsection{Network-Embedded Semi-Thermodynamic analysis (NEST)}
A relatively straight-forward way to measure the effect of semi-thermodynamic constraints on the solution space, is to use an approach derived from NET analysis.

\begin{eqnarray}
\textsf{\textbf{NEST}} && \nonumber\\
\forall k && \mathrm{\min/\max} ~~\Delta_r G'_k\nonumber\\
\textsf{such that:} && \nonumber\\
\forall  v_i > 0 && \Delta_r G'_i \leq -\varepsilon \\
\forall  v_i < 0 && \Delta_r G'_i \geq \varepsilon \\
\mathbf{\Delta_r G'} &=& \Sint ^\top \cdot \mathbf{\Delta_f G'} \\
\mathbf{\Delta_f G'} &\in& \mathcal{R}^{m}\\
\forall i~\textsf{in Table \ref{table:potentials}}\nonumber\\
\Delta_f G'_i &\geq& \Delta_f G'^\circ_i + RT \cdot \ln(b^L_i) \\
\Delta_f G'_i &\leq& \Delta_f G'^\circ_i + RT \cdot \ln(b^U_i) 
\end{eqnarray}
Just as in st-FBA, the thermodynamic constraints are only applied to a subset of metabolites (i.e. energy co-factors which we know with high confidence).

\subsection{Implementation of st-FBA}
The semi-thermodynamic Flux Balance Analysis algorithm was implemented using COBRApy \cite{Ebrahim2013-vw} and can be found at \url{https://github.com/eladnoor/stFBA}.

\bibliography{stfba_lib}
\bibliographystyle{plain}

\end{document}